\documentclass[prd,twocolumn,showpacs,floatfix,amsmath,nofootinbib,amssymb,floatfix]{revtex4}
\usepackage{graphicx,color,dcolumn,booktabs,bm,multirow}
\usepackage{longtable,lscape}
\usepackage{txfonts}
\usepackage{overpic}
\usepackage{amssymb}
\usepackage{indentfirst}
\usepackage{feynmf}   
\usepackage{slashed}  
\usepackage{cases}
\usepackage{color}
\usepackage{multirow}
\usepackage{epstopdf}
\usepackage{graphicx,color,dcolumn,booktabs,bm}

\usepackage[colorlinks,
            citecolor=blue,
            anchorcolor=red,
            menucolor=red,
            linkcolor=red,
            filecolor=red,
            runcolor=red,
            urlcolor=blue,
            frenchlinks=red]{hyperref}

\def\lrpartial{\buildrel\leftrightarrow\over\partial}
\begin{document}

\title{Radiative decays of the neutral $Z_c(3900)$}
\author{Dian-Yong Chen$^{1,2,3}$\footnote{Corresponding author}}\email{chendy@impcas.ac.cn}
\author{Yu-Bing Dong$^{4,5}$}\email{dongyb@ihep.ac.cn}
\affiliation{$^1$Institute of Modern Physics, Chinese Academy of Sciences, Lanzhou 730000, People's Republic of China\\
$^2$Research Center for Hadron and CSR Physics, Lanzhou
University $\&$ Institute of Modern Physics of CAS, Lanzhou 730000,
 People's Republic of China\\
$^3$State Key Laboratory of Theoretical Physics, Institute of Theoretical Physics, Chinese Academy of Sciences, Beijing 100190, People's Republic of China\\
$^4$ Institute of High Energy Physics, Chinese Academy of Sciences, Beijing 100049, People's Republic of China\\
$^5$ Theoretical Physics Center for Science Facilities (TPCSF), Chinese Academy of Sciences, Beijing 100049, People's Republic of China}

\date{\today}
\begin{abstract}
We study the radiative decays of the $Z_c(3900)^0$ in a hadronic molecule picture, where the $Z_c(3900)$ is treated as a $D \bar{D}^\ast +c.c$ hadronic molecule. The partial widths of $\Gamma(Z_c(3900)^0 \to \gamma \eta_c(2S))$ and $\Gamma(Z_c(3900)^0 \to \gamma \chi_{c0})$ are predicted to be of order 10 keV and the cross sections for $\sigma(e^+ e^- \to \pi^0 Z_c(3900) \to \pi^0 \gamma \eta_c(2S))$ and $\sigma(e^+ e^- \to \pi^0 Z_c(3900) \to \pi^0 \gamma \chi_{c0})$ are of order 0.1 pb at 4.23 GeV, which may be accessible for the BESIII and forthcoming Belle II.
\end{abstract}

\maketitle

\section{Introduction}\label{sec1}
As the first confirmed charged charmoniumlike state, $Z_c(3900)$ was first reported by the BESIII \cite{Ablikim:2013mio} and Belle \cite{Liu:2013dau} Collaborations in the $J/\psi \pi^{\pm}$ invariant mass spectrum of $e^+ e^- \to \pi^+ \pi^- J/\psi $ at 4.26 GeV. The CLEO Collaboration confirmed the existence of the $Z_c(3900)^\pm$ in the same process but at $4.17$ GeV and the neutral $Z_c(3900)$ was also reported with a $3\sigma$ significance \cite{Xiao:2013iha}. From the first observed mode of $Z_c(3900)$, i.e. the $J/\psi \pi^{\pm}$ mode, one can find that the charmoniumlike state $Z_c(3900)$ contains at least four quarks. In addition, the reported mass of the $Z_c(3900)$ is close to the $D\bar{D}^\ast$ mass threshold, which indicates this state could be a good candidate of the $D^\ast \bar{D} +c.c$ hadronic molecule. These peculiar properties of $Z_c(3900)$ stimulate the theoreticians and experimentalists to further investigate this state.

From the experimental side, after the observation of the $Z_c(3900)$ in the $J/\psi \pi$ invariant mass spectrum, more decay modes were searched for. In the $h_c \pi^\pm$ invariant mass spectrum of $e^+ e^- \to h_c \pi^+ \pi^-$, no evident signal of the $Z_c(3900)$ is observed and the significance of $Z_c(3900)$ is reported to be $2.1\sigma$ \cite{Ablikim:2013wzq}. The neutral partner of $ Z_c(3900)^\pm$ was recently observed with a significance of $10.4\sigma$ in the $\pi^0 J/\psi$ invariant mass spectrum of $e^+ e^- \to \pi^0 \pi^0 J/\psi$ \cite{Ablikim:2015tbp}. The Born cross sections for the $e^+ e^- \to \pi^0 \pi^0 J/\psi$ in the energy range from 4.190 to 4.42 GeV are about half of those $e^+e^-\to \pi^+ \pi^- J/\psi$, which is consistent with the isospin symmetry expectation for resonances. Besides the hidden charm mode, the BESIII Collaboration also found  $Z_c(3900)$ in the open charm process, such as $e^+ e^- \to \pi^\pm (D \bar{D}^\ast)^{\mp}$ \cite{Ablikim:2013xfr}, where the observed mass and width are $2\sigma$ and $1\sigma$, respectively, below those observed in the $\pi^\pm J/\psi$ mass spectrum by BESIII \cite{Ablikim:2013mio} and Belle \cite{Liu:2013dau} Collaborations. Recently, the BESIII Collaboration \cite{Ablikim:2015swa} confirmed the observation of $Z_c(3900)$ in $e^+ e^- \to \pi^\pm (D\bar{D}^\ast)^\mp$ process with the double $D$ tag technique. And more recently, the BESIII Collaboration reported their observation of the neutral $Z_c(3900)$ in $e^+ e^- \to \pi^0 (D\bar{D}^\ast)^0$ \cite{Ablikim:2015gda}. The cross sections of $\sigma(e^+ e^- \to Z_c(3885)^0 \pi^0 , Z_c(3885)^0 \to (D\bar{D}^\ast)^0)$ at 4226 MeV and 4257 MeV are reported to be $ 77 \pm 13 \pm 17 $ pb and $ 49 \pm 9 \pm 10 $ pb, respectively. In addition, the Compass Collaboration \cite{Adolph:2014hba} performed the search for $Z_c(3900)$ from the exclusive photoproduction process and an upper limit for the ratio $\mathcal{B}(Z_c(3900) \to \pi J/\psi) \times \sigma(\gamma N \to Z_c(3900) N)/\sigma(\gamma N \to J/\psi N)$ of $3.7 \times 10^{-3}$ was reported at the $90\%$ confidence level.

Since the charged $Z_c(3900)$ contains at least four quarks, $Z_c(3900)$ could be a good candidate for the tetraquark state. Since $Z_c(3900)$ is very close to another longstanding charmouniumlike state $X(3872)$, it was considered as the charged partner of the $X(3872)$ in a tetraquark scenario \cite{Maiani:2004vq}. In Ref. \cite{Ali:2011ug}, the authors proposed two tetraquark states $Z_c(3752)$ and $Z_c(3885)$, where the latter one was close to the observed $Z_c(3900)$.
The decays and the mass of the tetraquark state corresponding to $Z_c(3900)$ were estimated in Refs. \cite{Dias:2013xfa, Wang:2013vex} and the spectra of tetrquark states in a potential model also supported $Z_c(3900)$ as a tetraquark state \cite{Deng:2014gqa}. Besides the tetraquark interpretation, $Z_c(3900)$ had been predicted as the charmonium analog of the $Z_b(10610)$ and $Z_b(10650)$ with the initial single pion emission (ISPE) mechanism \cite{Chen:2011xk, Chen:2012yr,Chen:2011pv}. In Ref. \cite{Chen:2013coa}, the invariant mass distributions of $ \pi J/\psi$ and $\pi^+ \pi^-$ of $Y(4260) \to \pi^+ \pi^- J/\psi$ were reproduced by considering the interferences between the ISPE mechanism and dipion resonance contributions. In Ref. \cite{Swanson:2014tra}, the observed structures of $Z_b$ and $Z_c$ were explained as pure coupled channel cusp effects other than resonances.

The observed mass of the $Z_c(3900)$ is very close to the threshold of $D^\ast \bar{D}$, which indicates that the $Z_c(3900)$ could be a $D^\ast \bar{D}$ hadronic molecule state. In Refs. \cite{Sun:2011uh, Sun:2012zzd, Aceti:2014uea}, the potential of $D^\ast \bar{D}$ had been evaluated and by solving the Schr$\mathrm{\ddot{o}}$dinger equation, and they found bound state solution for the $D^\ast \bar{D}$ system, which corresponded well to the observed $Z_c(3900)$. The QCD sum rule calculations in Refs. \cite{Wang:2013daa, Chen:2015ata} also supported that $Z_c(3900)$ could be a deuteronlike hadronic molecule state. In Ref. \cite{Wilbring:2013cha}, the electromagnetic structure of the $Z_c(3900)$ was investigated in a $D^\ast \bar{D}+c.c$ framework. The product and decay behaviors of $Z_c(3900)$ had been studied in a $D^\ast \bar{D}$ hadronic molecule scenario with the Weinberg compositeness condition in Refs. \cite{Dong:2013iqa, Dong:2013kta}, and the theoretical estimations were consistent with the corresponding experimental measurements. Besides the observed channels, some other decay modes of $Z_c(3900)$ have been studied in the molecule scenario, such as the $\rho \eta_c$, $J/\psi \pi \gamma$ \cite{Li:2014pfa, Gutsche:2014zda, Esposito:2014hsa, Ke:2013gia}. All these theoretical studies supported that $Z_c(3900)$ could be assigned as a $D^\ast \bar{D}$ hadronic molecule state.

In this way, we study the radiative decay of the neutral $Z_c(3900)$ in a $D^\ast \bar{D}$ hadronic molecule framework. The radiative transitions between $Z_c^0(3900)$ and the charmonia states are particular modes compared to the charged $Z_c(3900)$. The quark and antiquark in the different components of $Z_c^0(3900)$ can annihilate into a photon and the rest charm and anticharm quark form a charmonium in the final state. Considering the $J^{PC}$ quantum numbers conservation, $Z_c(3900)^0$ can radiatively transit into $\eta_c$, $\eta_c(2S)$, and $\chi_{cJ}$. In the present work, we will evaluate these radiative transitions in a molecule scenario with an effective Lagrangian approach. The results in present work can provide further information for the experimental search for the $Z_c(3900)^0$, and, on the other hand, the experimental measurements for these radiative decay processes could be a crucial test for the molecule interpretation of $Z_c(3900)$. At present, the cross section for $e^+ e^- \to \pi^0 Z_c(3900)^0 \to \pi^0 \pi^0 J/\psi$ has been reported by the BESIII Collaboration, thus, we can take this process as a scale to calculate the cross section for $e^+ e^- \to \pi^0 Z_c(3900)^0 \to \pi^0 \gamma (c\bar{c})$, where $(c\bar{c})$ denotes $\eta_c$, $\eta_c(2S)$, and $\chi_{cJ}$.

This work is organized as follows: The $J/\psi \pi^0$ and the radiative decays of the $Z_c(3900)^0$ in $D\bar{D}^\ast +\mathrm{c.c}$ molecule scenario are presented in the following section. The numerical results for the decays of $Z_c(3900)^0$ are presented in Section \ref{sec3}, and Section \ref{sec4} is dedicated to a short summary.

\section{$J/\psi \pi^0$ and the radiative decays of $Z_c^0(3900)^0$} \label{sec2}
In the $D\bar{D}^\ast +c.c$ molecule scenario, the charmoniumlike state $Z_c(3900)^0$ couples to its components dominantly via an $S$ wave. The simplest effective Lagrangian describing the coupling of $Z_c(3900)^0$ and its components is in the form,
\begin{eqnarray}
\mathcal{L}_{Z_c D^\ast D}  &=& \frac{g_{Z_c}}{2} Z_c(x) \int dy \Phi(y) \Big[\Big(D^0(x+\omega_{D^\ast} y) \bar{D}^{\ast0}(x-\omega_D y) \nonumber\\&& \quad-D^+(x+\omega_{D^\ast} y) D^{\ast-}(x-\omega_D y) \Big) + \mathrm{H.C.} \big) \Big] \label{Eq:Zc}
\end{eqnarray}
where $\omega_{D^{(\ast)}} =m_{D^{(\ast)}}/(m_{D}+ m_{D^\ast})$. The effective correlation function $\Phi(y)$ is introduced to describe the distribution of $D^{(\ast)}/\bar{D}^{(\ast)}$ in the hadronic molecule $Z_c(3900)^0$ and also plays the role to render the Feynman diagrams ultraviolet finite, which indicates the Fourier transform of $\Phi(y)$ should vanish sufficiently fast in the ultraviolet region of the Euclidean space. In the present work, we adopt a Gaussian form for the correlation function, which has been widely used to study the hadronic molecule decays \cite{Dong:2014ksa, Dong:2013iqa, Dong:2008rad, Dong:2013kta, Dong:2012hc, Dong:2009yp, Faessler:2008zza, Faessler:2007gv}. The Fourier transform of the correlation function $\tilde{\Phi}(y)$ is in the form,
\begin{eqnarray}
\tilde{\Phi}(P_E) \doteq \mathrm{exp}(-P_E^2/\Lambda^2),\label{Eq:Corr}
\end{eqnarray}
where $P_E$ is the Jacobi momentum in Euclidean space and $\Lambda$ is a parameter which characterizes the distribution of components in the hadronic molecule. The value of $\Lambda$ is of order 1 GeV and dependent on a different system \cite{Dong:2014ksa, Dong:2013iqa, Dong:2013kta, Dong:2012hc, Dong:2009yp, Faessler:2008zza, Faessler:2007gv}.

\begin{figure}[htb]
\centering
\includegraphics[width=85mm]{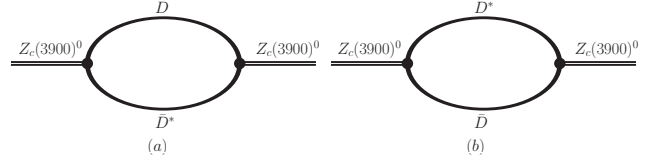}
\caption{The mass operator of $Z_c(3900)^0$. }\label{Fig:mo}
\end{figure}

The coupling of the hadronic molecule $Z_c(3900)^0$ to its components $D\bar{D}^\ast + c.c$ can be estimated by the compositeness condition \cite{Weinberg:1962hj, Salam:1962ap}. For a composite particle, the renormalization constant of the particle wave function is zero, i.e.,
\begin{eqnarray}
Z_{Z_c} =1-\left.\frac{d\Sigma_{Z_c}(s)}{ds}\right|_{ s=m_{Z_c}^2} \equiv 0,
\end{eqnarray}
where $\Sigma_{Z_c}$ is the transverse part of the mass operator $\Sigma_{Z_c}^{\mu \nu}$ of the hadronic molecule $Z_c(3900)^0$, which is is defined as,
\begin{eqnarray}
\Sigma^{\mu \nu}_{Z_c}(p) =g_\perp^{\mu \nu}\Sigma_{Z_c}(p) +\frac{p^\mu p^\nu}{p^2} \Sigma_{Z_c}^{L}(p)
\end{eqnarray}
with $g_{\perp}^{\mu \nu}=g^{\mu \nu} -p^\mu p^\nu/p^2$. From the mass operator presented in Fig. \ref{Fig:mo}, we can get the  concrete form of $\Sigma^{\mu \nu}_{Z_c}$ as
\begin{eqnarray}
\Sigma^{\mu \nu}_{Z_c}(p)&=&g_{Z_c}^2 \int \frac{d^4q}{(2 \pi)^4} \tilde{\Phi}^2(q-\omega_D p) \frac{-g^{\mu \nu} +q^\mu q^\nu/m_{D^\ast}^2}{q^2 -m_{D^\ast}^2}\nonumber\\&&\times \frac{1}{(p-q)^2-m_D^2}, \nonumber
\end{eqnarray}
where $p^2=m_{Z_c}^2$ and in present calculation, we set $m_{Z_c}= m_D+m_{D^\ast}-E_b$ with $E_b$ to be the binding energy of $Z_c(3900)^0$.

\subsection{The decay of $Z_c(3900)^{0} \to \pi^0 J/\psi$}
\begin{figure}[htb]
\centering
\includegraphics[width=80mm]{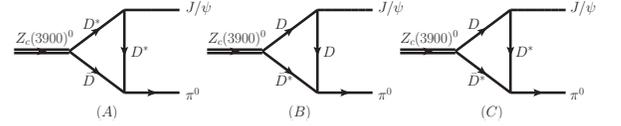}
\caption{Diagrams contributing to the decay $Z_c(3900)^0 \to \pi^0 J/\psi$. }\label{Fig:feyn-jpsipi}
\end{figure}

In the hadronic molecule picture, $Z_c(3900)$ can decay into $J/\psi \pi$ by rearranging the quarks in its components. At the hadron level, $Z_c(3900)$ is treated as a bound state of $D\bar{D}^\ast +c.c$ and the decay $Z_c(3900) \to \pi J/\psi$  occurs by exchanging a proper charmed meson as shown in Fig. \ref{Fig:feyn-jpsipi}. In the present work, we estimate these triangle diagrams in an effective Lagrangian approach. Besides the Lagrangian present in Eq. (\ref{Eq:Zc}), the couplings of the charmed mesons with pion and $J/\psi$ are needed, which can be constructed by the heavy quark limit and chiral symmetry. The concrete form of the involved Lagrangians is \cite{Oh:2000qr,Casalbuoni:1996pg,Colangelo:2002mj},
\begin{eqnarray}
\mathcal{L}_{\psi D^{(\ast)} D^{(\ast)}}
&=& -ig_{\psi DD } \psi_\mu (\partial^\mu
D D^\dagger- D
\partial^\mu D^\dagger) \nonumber\\
&& + g_{\psi
D^\ast D} \varepsilon^{\mu \nu \alpha \beta}
\partial_\mu \psi_\nu (D^\ast_\alpha \lrpartial_\beta
D^\dagger -D \lrpartial_\beta
D_\alpha^{\ast \dagger} ) \nonumber\\
&& + ig_{\psi
D^\ast D^\ast} \psi^\mu
(D^\ast_\nu \partial^\nu D^{\ast \dagger}_\mu
-\partial^\nu D^{\ast}_\mu D^{\ast \dagger}_\nu
\nonumber\\&&-D^\ast_\nu \lrpartial_\mu D^{\ast \nu
\dagger}),\nonumber\\
\mathcal{L}_{D^{(\ast)} D \pi}&=&  ig_{D^\ast D
\pi} (D^\ast_{\mu} \partial^\mu \pi {D}^\dagger-D \partial^\mu \pi
\bar{D}^{\ast \dagger}_{\mu})\nonumber\\&&-g_{D^\ast D^\ast \pi} \varepsilon^{\mu \nu
\alpha \beta}
D^\ast_{\mu}\partial_{\nu} \pi \partial_{\alpha}
\bar{D}^{\ast\dagger}_{\beta},
\label{Eq:Lag1}
\end{eqnarray}
where the charm meson isodoublets can be defined as $D^{(*)}=(D^{(*)0},D^{(*)+})$ and $\pi={\mbox{\boldmath $\tau$}}\cdot {\mbox{\boldmath $\pi$}}$ \cite{Oh:2000qr}.

In the heavy quark limit, the coupling constants among the charmonia and
charmed mesons satisfy \cite{Casalbuoni:1996pg, Colangelo:2002mj},
\begin{eqnarray}
g_{\psi D D} &=& 2g_2 \sqrt{m_{\psi}} m_{D},\nonumber\\
g_{\psi D^\ast D} &=& 2g_2  \sqrt{m_{D} m_{D^\ast}/m_{\psi}},\nonumber\\
g_{\psi D^\ast D^\ast} &=& 2g_2 \sqrt{m_{\psi}} m_{D^\ast},
\end{eqnarray}
where $g_2= \sqrt{m_{\psi}}/(2 m_D f_{\psi})$ is a gauge coupling with $f_{\psi}$ to be the decay constant of a charmonium $\psi$. With the center values of the leptonic partial decay widths, i.e., $\Gamma({J/\psi\to e^+e^-})=5.55 \pm 0.14 \pm0.02\ \mathrm{keV}$  and $\Gamma({\psi(2S)\to e^+e^-})= 2.36 \pm 0.04 \ \mathrm{keV} $\cite{Agashe:2014kda}, one can obtain $f_{J/\psi}=416\ \mathrm{MeV}$ and $f_{\psi(2S)} =296\ \mathrm{MeV}$. Considering the chiral symmetry and the heavy quark limit, the coupling constants among the charmed mesons and pion can be related to the gauge coupling $g$ by the relations,
\begin{eqnarray}
g_{D^\ast D^\ast \pi}=g_{D^\ast D \pi}/\sqrt{m_D m_{D^\ast}}
=2g/f_\pi,\label{couple3}
\end{eqnarray}
where  $f_\pi=132\ \mathrm{MeV}$ is the pion decay constant and $g=0.59$ is extracted from the experimental measurement of the partial decay width of $D^\ast \to D \pi$ \cite{Agashe:2014kda}.

With the above preparations, we can obtain the amplitudes for $Z_c(3900)^0(p_0)\rightarrowtail [D^{(\ast)}(p_1) \bar{D}^{(\ast)}(p_2) ]D^{(\ast)}(q) \rightarrowtail J/\psi(p_3) \pi^0(p_4)$, which correspond to the diagrams in Fig. \ref{Fig:feyn-jpsipi}. The concrete forms of the amplitudes are,
\begin{eqnarray}
\mathcal{M}_A &=& \int \frac{d^4q}{(2 \pi)^4}\Big[\frac{g_{Z_c}}{2} \epsilon^{\mu}_{Z_c} \tilde{\Phi}(\omega_{D} p_1 -\omega_{D^\ast} p_2)\Big] \nonumber\\ &&\times \Big[ig_{\psi D^\ast D^\ast} \epsilon_{\psi}^\nu (iq_\rho g_{\nu \lambda}+ip_{1\lambda} g_{\rho \nu}- (iq_\nu+ip_{1\nu})g_{\rho \lambda})\Big]\nonumber\\&&\times [-g_{D^\ast D \pi} (ip_{4\alpha}) ] \frac{-g_{\mu}^{\rho}+p_{1\mu} p_1^\rho/m_{D^\ast}^2}{p_1^2-m_{D^\ast}^2} \frac{1}{p_2^2-m_{D}^2}\nonumber\\&&\times \frac{-g^{\lambda \alpha}+q^\lambda q^\alpha/m_{D^\ast}^2}{q^2-m_{D^\ast}^2},\nonumber \\
\mathcal{M}_B &=&\int \frac{d^4q}{(2 \pi)^4}\Big[\frac{g_{Z_c}}{2} \epsilon^{\mu}_{Z_c} \tilde{\Phi}(\omega_{D} p_1 -\omega_{D^\ast} p_2)\Big]  \nonumber\\&&\times \Big[-ig_{\psi DD} \epsilon_{\psi}^\nu (-ip_{1\nu}-iq_{\nu})\Big] \Big[ ig_{D^\ast D\pi} (-ip_{4\alpha})\Big] ,\nonumber \\&&\times \frac{1}{p_1^2 -m_D^2} \frac{-g_{\mu}^\alpha+p_{2\mu} p_2^\alpha/m_{D^\ast}^2}{p_2^2 -m_{D^\ast}^2} \frac{1}{q^2-m_D^2}\nonumber\\
\mathcal{M}_C &=& \int \frac{d^4q}{(2 \pi)^4}\Big[\frac{g_{Z_c}}{2} \epsilon^{\mu}_{Z_c} \tilde{\Phi}(\omega_{D} p_1 -\omega_{D^\ast} p_2)\Big] \nonumber\\ &&\times \Big[g_{\psi D^\ast D} \varepsilon_{\rho \nu \lambda \phi} (ip_3^\rho)\epsilon_{\psi}^\nu (-iq^\phi +ip_1^\phi)\Big]\nonumber\\&&\times \Big[-g_{D^\ast D^\ast \pi} \varepsilon_{\alpha \beta \kappa \delta} (ip_4^\beta) (-ip_2^\kappa) \Big] \frac{1}{p_1^2-m_D^2} \nonumber\\ &&\times \frac{-g_\nu^\alpha+ p_{2\nu}p_2^\alpha/m_{D^\ast}^2}{p_2^2-m_{D^\ast}^2} \frac{-g_\nu^\delta +q_\nu q^\delta/m_{D^\ast}^2}{q^2-m_{D^\ast}^2} .
\end{eqnarray}

The total amplitude of $Z_c(3900)^0 \to J/\psi \pi^0$ is,
\begin{eqnarray}
\mathcal{M}_{Z_c(3900)^0 \to \pi^0 J/\psi}^{\mathrm{Tot}} =4 (\mathcal{M}_A+ \mathcal{M}_B+ \mathcal{M}_C) ,
\end{eqnarray}
where the factor $4$ comes from the isospin and charge symmetry. After performing the loop integral in the amplitudes, the total amplitude of $Z_c(3900)^0(p_0) \to J/\psi(p_3) \pi^0(p_4)$ can reduce to the form,
\begin{eqnarray}
\mathcal{M}_{Z_c(3900)^0\to \pi^0 J/\psi }^{\mathrm{Tot}}=\epsilon_{Z_c}^\mu \epsilon_{\psi}^\nu (g_S g_{\mu \nu} +g_D p_{3 \mu} p_{4 \nu}) \label{Eq:jpsipi},
\end{eqnarray}
where the coupling constants $g_S$ and $g_D$ can be estimated from the loop integral of the amplitudes.

\subsection{The radiative decay of $Z_c(3900)^0$}
For the neutral $Z_c(3900)$, the light quarks in the meson components are the same; thus, the radiative transition to the conventional charmonium can occur via annihilating the light quarks. In the present work, we consider $Z_c(3900)^0$ radiatively decaying into $\eta_c(1S)/\eta_c(2S)$, $\chi_{c0}$ and $\chi_{c1}$ in an effective Lagrangian approach. The involved effective Lagrangians of charmonium and charmed mesons are \cite{Oh:2000qr,Casalbuoni:1996pg,Colangelo:2002mj} 
\begin{eqnarray}
\mathcal{L}_{\eta_c D^{(\ast)}  D^{(\ast)}}&=& i g_{\eta_c D^\ast D}
D^{\ast \mu}(\partial_\mu \eta_c {D}^\dagger-\eta_c
\partial_\mu  {D}^\dagger) +H.C. \nonumber\\& -& g_{\eta_c D^\ast D^\ast}
\varepsilon^{\mu\nu\alpha\beta}
\partial_\mu D^\ast_\nu {D}^{\ast \dagger}_\alpha \partial_\beta\eta_c,\nonumber\\
\mathcal{L}_{\chi_{cJ} D^{(\ast)}  D^{(\ast)}}&=& - g_{\chi_{c0} D D } \chi_{c0} D
D^\dagger - g_{\chi_{c0} D^\ast
D^\ast} \chi_{c0} D_{\mu}^\ast D^{\ast
\mu\dagger } \nonumber\\ &&  +i g_{\chi_{c1} D
D^\ast} \chi_{c1}^\mu ( D^{\ast }_\mu
D^\dagger - D D^{\ast \dagger}_\mu ),\label{Eq:Lag2}
\end{eqnarray}
where the coupling $g_{\eta_c D^\ast D }$ and $g_{\eta_c D^\ast D^\ast}$ can be related to the gauge coupling $g_2$ by \cite{Casalbuoni:1996pg},
\begin{eqnarray}
g_{\eta_c D^\ast D} &=& 2g_2 \sqrt{m_{\eta_c} m_D m_{D^\ast}}\nonumber\\
g_{\eta_c D^\ast D^\ast} &=& 2g_2 m_{D^\ast} /\sqrt{m_{\eta_c}}
\end{eqnarray}
and the coupling related to $\eta_c(2S)$ can be evaluated with the above relation by replacing the mass of $\eta_c$ with $\eta_c(2S)$ and the mass and decay constants of $J/\psi$ in $g_2$ with the corresponding values for $\psi(2S)$. As for the coupling constants related to $\chi_{c0}$ and $\chi_{c1}$, in the heavy quark limit, they can be related to the gauge coupling $g_1$ by,
\begin{eqnarray}
g_{\chi_{c0}DD} &=&2 \sqrt{3} g_1 \sqrt{m_{\chi_{c0}}} m_D ,\nonumber\\
g_{\chi_{c0}D^\ast D^\ast} &=&\frac{2}{\sqrt{3}} g_1 \sqrt{m_{\chi_{c0}}} m_{D^\ast} ,\nonumber\\
g_{\chi_{c1} D^\ast D}  &=& 2\sqrt{2} g_1 \sqrt{m_{\chi_{c1}} m_D m_{D^\ast}},
\end{eqnarray}
where $g_1 =-\sqrt{m_{\chi_{c0}}/3}/f_{\chi_{c0}}$ and $f_{\chi_{c0}}= 0.51$ GeV is the decay constant of $\chi_{c0}$ \cite{Colangelo:2002mj}.

For the discussed radiative decay process, we still need the effective Lagrangians for $\gamma DD$ and $\gamma D^\ast D^\ast$, which can be constructed from the Lagrangians for the free scalar and massive vector fields by minimal substitution $\partial^\mu \to \partial^\mu +ieA^\mu$, and the concrete form of the Lagrangians is \cite{Chen:2010re},
\begin{eqnarray}
\mathcal{L}_{{DD}\gamma} &=& i e A_{\mu} D^{-} \lrpartial^{\mu}
D^{+}, \nonumber\\
{L}_{{D}^{*} {D}^{*} \gamma} &=& i e A_{\mu} \Big\{ g^{\alpha
\beta} D_{\alpha}^{*-} \lrpartial^{\mu} D_{\beta}^{*+} + g^{\mu
\beta} D_{\alpha}^{*-} \partial^{\alpha} D_{\beta}^{*+}\nonumber\\&& - g^{\mu
\alpha } \partial^{\beta} D_{\alpha}^{*-}  D_{\beta}^{*+}\Big\},
\label{Eq:Lag3}
\end{eqnarray}
where the neutral interactions vanish. The interaction of $D^\ast D \gamma $ is in the form,
\begin{eqnarray}
\mathcal{L}_{\mathcal{D}^{*} \mathcal{D} \gamma} &=& \bigg\{\frac{e}{4}
g_{{D^{*+}D^+}\gamma} \varepsilon^{\mu \nu \alpha \beta} F_{\mu
\nu} {D}^{*+}_{\alpha \beta} {D}^+ \nonumber\\&&+\frac{e}{4}
g_{{D^{*0}{D}^0}\gamma} \varepsilon^{\mu \nu \alpha \beta} F_{\mu
\nu} \mathcal{D}^{*0}_{\alpha \beta} {{D}}^0\bigg\}+ h.c \  . \label{Eq:Lag4}
\end{eqnarray}
where $F_{\mu \nu} =\partial_{\mu} A_{\nu} - \partial_{\nu} A_{\mu}$
and $D^{*0,+}_{\mu \nu} = \partial_{\mu} D^{*0,+}_{\nu} -
\partial_{\nu} D^{*0,+}_{\mu}$ are the stress tensors of the photon and vector charmed meson, respectively. The introduced coupling constants $g_{D^\ast D\gamma}$ can be evaluated by the corresponding partial decay width of $D^\ast \to D\gamma$. According to the center values of the PDG average for the branching ratio of $D^{\ast+ } \to D^+ \gamma$ and the total width of $D^{\ast+}$, we can roughly estimate $g_{D^{\ast+} D^+ \gamma} \simeq 0.5\  \mathrm{GeV}$. However, as for the neutral case,  the $D^{\ast 0}$ total width is kept unknown. Considering the isospin symmetry of the strong decay processes, we can relates the partial width of $D^{\ast0} \to D^0 \pi^0$ to the one of $D^{\ast+} \to D^+ \pi^0$ by  \cite{Dong:2008gb},
\begin{eqnarray}
\Gamma(D^{\ast 0} \to D^0 \pi^0) \simeq \frac{\mathrm{PS}_{D^{\ast +} \to D^+ \pi^0}}{\mathrm{PS}_{D^{\ast 0} \to D^0 \pi^0}} \Gamma(D^{\ast +} \to D^+ \pi^0),
\end{eqnarray}
where $\mathrm{PS}_{D^{\ast } \to D \pi^0}$ is the phase space of $D^\ast \to D \pi^0$. The phase space factor is introduced due to the small phase spaces of the decay processes. With this relation, we can evaluate the partial width of $D^{\ast 0} \to D^0 \pi^0$. Together with the ratio of $\Gamma(D^{\ast 0} \to D^0 \pi^0)$ and $\Gamma(D^{\ast 0} \to D^0 \gamma^0)$, $\Gamma(D^0 \pi^0)/\Gamma(D^0 \gamma) =1.74 \pm 0.02 \pm 0.13$ \cite{Agashe:2014kda}, one can estimate the partial decay width of $D^{\ast 0} \to D^0 \gamma$ and then the coupling constant $g_{D^{\ast 0} D^0 \gamma}$. Here, we adopt the center value of the partial width ratio and obtain $g_{D^{\ast 0} D^0 \gamma}\simeq 2.0\ \mathrm{GeV}$.

\begin{figure}[htb]
\centering
\includegraphics[width=85mm]{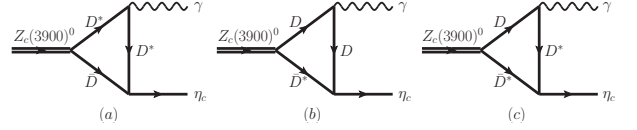}
\caption{Diagrams contributing to the radiative decay $Z_c(3900) \to \eta_c \gamma$. The diagrams related to $\eta_c(2S)$ can be obtained by replacing $\eta_c$ with $\eta_c(2S)$ in the diagrams. }\label{Fig:feyn-etac}
\end{figure}

\begin{figure}[htb]
\centering
\includegraphics[width=55mm]{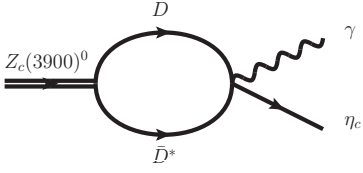}
\caption{Contact diagrams contributing to the radiative decay $Z_c(3900) \to \eta_c \gamma$. }\label{Fig:feyn-con}
\end{figure}

With the above effective Lagrangian, we can construct the amplitudes of $Z_c(3900)^0(p_0) \to \gamma(p_3) \eta_c(p_4)$, which correspond to the diagrams in Fig. \ref{Fig:feyn-etac}. After loop integral, the amplitudes can be reduced as,
\begin{eqnarray}
\mathcal{M}_{Z_c(3900)^0 \to \gamma \eta_c(1S)}^{\mathrm{Tri}} =\epsilon_{Z_c}^\mu \epsilon_{\gamma}^\nu (g_S^{\mathrm{Tri}} g_{\mu \nu} +g_D^{\mathrm{Tri}} \frac{p_3^\mu p_4^\nu}{p_3\cdot p_4}).
\end{eqnarray}
One should notice that this amplitude satisfies the gauge invariance of the photon field only when $g_S^{\mathrm{Tri}}= -g_D^{\mathrm{Tri}}$, but actually, this equation does not hold if only triangle diagrams are included. From the interaction of $\eta_c$ and $D^\ast D$ in Eq. (\ref{Eq:Lag2}), one can construct a four point interaction $\eta_c D^\ast D\gamma$ by minimal substitution, which is,
\begin{eqnarray}
\mathcal{L}_{\eta_c D^{\ast} D\gamma} &=&  g_{\eta_c
D^\ast D \gamma} D^{\ast \mu} \eta_c A_\mu D^\dagger +H.C.
\end{eqnarray}
With this four point effective interaction, we find that the contact diagram in Fig. \ref{Fig:feyn-con} will also contribute to the radiative decay $Z_c^{0}(3900) \to \gamma \eta_c$. The amplitude of the contact diagram is in the form,
\begin{eqnarray}
&&\mathcal{M}^{\mathrm{Con}}_{Z_c(3900)^0 \to \gamma \eta_c} = \int \frac{d^4q}{(2 \pi)^4}\Big[\frac{g_{Z_c}}{2} \epsilon^{\mu}_{Z_c} \tilde{\Phi}(\omega_{D} p_1 -\omega_{D^\ast} p_2)\Big]\nonumber\\
&&\qquad \times\Big[g_{\eta_c D^\ast D \gamma} \epsilon_{\gamma}^\nu\Big] \frac{1}{(p_0-q)^2-m_D^2} \frac{-g_{\mu \nu} +q_\mu q_\nu/m_{D^\ast}^2}{q^2 -m_{D^\ast}^2},
\end{eqnarray}
which can be further reduced into the form \cite{Chen:2013cpa},
\begin{eqnarray}
\mathcal{M}^{\mathrm{Con}}_{Z_c(3900)^0 \to \gamma \eta_c}= g_{S}^{\mathrm{Con}} \epsilon_{Z_c}^\mu \epsilon_\gamma^\nu g_{\mu \nu}. \label{Eq:Con}
\end{eqnarray}

Then the total amplitude of $Z_c(3900)^0 \to \gamma \eta_c$ is the sum of the contributions from the triangle diagrams and contact diagram, i.e.,
\begin{eqnarray}
\mathcal{M}_{Z_c(3900)^0 \to \gamma \eta_c}^{\mathrm{Tot}} &=& \mathcal{M}_{Z_c(3900)^0 \to \gamma \eta_c}^{\mathrm{Tri}}+ \mathcal{M}_{Z_c(3900)^0 \to \gamma \eta_c}^{\mathrm{Con}}\nonumber\\
&=& \epsilon_{Z_c}^\mu \epsilon_{\gamma}^\nu \Big[\big(g_S^{\mathrm{Tri}}+ g_S^{\mathrm{Con}}\big) g_{\mu \nu} +g_D^{\mathrm{Tri}} {p_3^\mu p_4^\nu\over p_3\cdot p_4}\Big], \nonumber\\
\end{eqnarray}
which should be satisfy with the gauge invariance of the photon field and $g_S^{\mathrm{Tri}} +g_S^{\mathrm{Con}}= -g_D^{\mathrm{Tri}}$. In our calculations, we find that the $g_S^{\mathrm{Tri}} +g_S^{\mathrm{Con}}$ is only approximately equal to the $-g_D^{\mathrm{Tri}}$. This small discrepancy could come from the contributions of the triangle diagrams with highly excite state, such as $D(2S),\ D^\ast(2S)$, as exchanged mesons. Fortunately, as present in Eq. (\ref{Eq:Con}), the contact diagram only contribute to the $S-$ wave term; thus, we can estimate the dominate contributions from the triangle diagrams and impose the condition $g_{S}^{\mathrm{Tri}} + g_{S}^{\mathrm{Con}}=-g_{D}^{\mathrm{Tri}}$ to keep the gauge invariance of the total amplitude of $Z_c^0(3900) \to \gamma \eta_c$, and the amplitude can be further expressed as,

\begin{eqnarray}
\mathcal{M}_{Z_c(3900)^0 \to \gamma \eta_c}^{\mathrm{Tot}} = g_D^{\mathrm{Tri}} \epsilon_{Z_c}^\mu \epsilon_{\gamma}^\nu \Big[ g_{\mu \nu} -{p_3^\mu p_4^\nu\over p_3\cdot p_4}\Big] \label{Eq:etac},
\end{eqnarray}
where the coupling constants $g_D^{\mathrm{Tri}}$ can be evaluated form loop integral of the amplitudes related to the triangle diagrams in Fig. \ref{Fig:feyn-etac}.

\begin{figure}[htb]
  \centering
  \includegraphics[width=70mm]{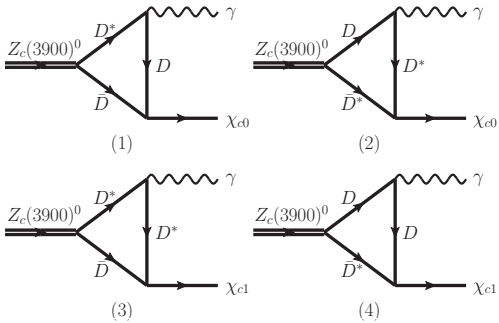}
  \caption{Diagrams contributing to the radiative decays $Z_c(3900) \to \chi_{cJ} \gamma\ (J=0,1)$. }\label{Fig:feyn-chicj}
\end{figure}

The possible diagrams contributing to $Z_c(3900)^0 \to \gamma \chi_{cJ},\ (J=0,1)$ are presented in Fig. \ref{Fig:feyn-chicj}. With the effective Lagrangians list in Eqs. (\ref{Eq:Lag2}) -(\ref{Eq:Lag4}), one can construct the amplitudes corresponding to Figs. \ref{Fig:feyn-chicj}(a)-\ref{Fig:feyn-chicj}(d). As for the $Z_c(3900)^0(p_0) \to \gamma(p_3) \chi_{c0}(p_4)$ process, the amplitudes can be reduced into the form,
\begin{eqnarray}
\mathcal{M}_{Z_c(3900)^0 \to \gamma \chi_{c0}}^{\mathrm{Tot}} =g_{Zc \chi_{c0} \gamma} \varepsilon_{\mu \nu \alpha \beta} \epsilon_{Z_c}^\mu \epsilon_\gamma^\nu p_3^\alpha p_4^\beta, \label{Eq:chic0}
\end{eqnarray}
which is gauge invariant for the photon fields and the coupling constants $g_{Zc \chi_{c0} \gamma}$ can be estimated by the loop integral of the amplitudes corresponding to Figs. \ref{Fig:feyn-chicj}(a) and \ref{Fig:feyn-chicj}(b).

As for the $Z_c(3900)^0 \to \gamma \chi_{c1}$, the amplitudes corresponding to Figs \ref{Fig:feyn-chicj}(c) and \ref{Fig:feyn-chicj}(d) can be reduced into the form
\begin{eqnarray}
\mathcal{M}_{Z_c(3900)^0\to \gamma \chi_{c1}}^{\mathrm{Tri}} &=&\epsilon_{Z_c}^\mu \epsilon_{\gamma}^\nu \epsilon_{\chi_{c1}}^\alpha \big(f_1^{\mathrm{Tri}} p_{3\alpha} g_{\mu \nu} +f_2^{\mathrm{Tri}} p_{3\mu} g_{\nu \alpha}\nonumber\\&& +f_3 p_{0\nu} g_{\alpha \mu}\big). \label{Eq:chic1a}
\end{eqnarray}
In the $Z_c(3900)^0$ rest frame, we have $p_0=\{m_{Z_c},\vec{0}\}$, while the zero component of the photon polarization vector is zero, i.e., $\epsilon_{\gamma}^0=0$, thus, $\epsilon_\gamma^\nu p_{0\nu}=\epsilon_{\gamma}^0 p_0^0-\vec{\epsilon}_\gamma \vec{p}_0\equiv 0$. With this equation, the term related to $f_3$ vanishes. In addition, $Z_c(3900)^0$ decays into $\gamma \chi_{c1}$ via the $P$ wave and $F$ wave, but in Eq. (\ref{Eq:chic1a}), $Z_c(3900)^0$ couples to $\gamma \chi_{c1}$ only via the $P$ wave, which leads to the amplitudes not being gauge invariant. The reason is that in the coupling of $Z_c D^\ast D$ and $\chi_{c1} D^\ast D$, only the lowest partial wave, the $S$ wave, is considered, but actually, $Z_c(3900)$ and $\chi_{c1}$ can also couple to $D^\ast D$ via the $D$ wave. If the higher partial wave couplings are considered in the effective Lagrangians in Eqs. (\ref{Eq:Lag1}) and (\ref{Eq:Lag3}), an additional term proportional to $p_{3}^\mu p_4^\nu p_3^\alpha$ will appear in the reduced amplitude. Then, the reduced amplitude of $Z_c(3900)^0 \to \gamma \chi_{c1}$ should be,
\begin{eqnarray}
\mathcal{M}_{Z_c(3900)^0\to \gamma \chi_{c1}} &=& \epsilon_{Z_c}^\mu \epsilon_{\gamma}^\nu \epsilon_{\chi_{c1}}^\alpha \big(f_1^\prime p_{3\alpha} g_{\mu \nu} +f_2^\prime p_{3\mu} g_{\nu \alpha}\nonumber\\&& + g_1 \frac{p_{3\mu} p_{4\nu } p_{3\alpha}}{p_3 \cdot p_4} \big). \label{Eq:chic1b}
\end{eqnarray}
Comparing to the $S$ wave coupling, the $D$ wave coupling should be suppressed. In addition, the higher partial wave coupling will be important in the large momentum region, but the correlation function in the loop integral will be further suppressed in this region. Thus, we suppose, the $D$ wave coupling in the vertexes  $Z_c(3900) D^\ast D $ and $\chi_{c1}D^\ast D$ will only contribute to the term $ g_1 p_{3\mu} p_4^\nu p_3^\alpha $ and $f_{1,2}^\prime\simeq f_{1,2}^{\mathrm{Tri}}$. To satisfy the gauge invariance of the photon field, the coupling constants in Eq. (\ref{Eq:chic1b}) should satisfy $g_1=-(f_1^\prime  +f_2^\prime)\simeq -(f_1^\mathrm{Tri}+f_2^\mathrm{Tri})$. In our calculation, we also find $f_1^{\mathrm{Tri}}$ and $f_2^{\mathrm{Tri}}$ are in different signs and $(f_1^\mathrm{Tri}+ f_2^\mathrm{Tri})$ is much smaller than the absolute values of $f_1^{\mathrm{Tri}}$ and $f_2^{\mathrm{Tri}}$, which is consistent with our analysis. With these assumptions, we have the following gauge invariant amplitude,
\begin{eqnarray}
\mathcal{M}_{Z_c(3900)^0 \to \gamma \chi_{c1}}^{\mathrm{Tot}} &=&\epsilon_{Z_c}^\mu \epsilon_{\gamma}^\nu \epsilon_{\chi_{c1}}^\alpha \big(f_1^{\mathrm{Tri}} p_{3\alpha} g_{\mu \nu} +f_2^{\mathrm{Tri}} p_{3\mu} g_{\nu \alpha}\nonumber\\&& - (f_1^{\mathrm{Tri}}+f_2^{\mathrm{Tri}}) \frac{p_{3\mu} p_{4\nu } p_{3\alpha}}{p_3 \cdot p_4} \big). \label{Eq:chic1c}
\end{eqnarray}

With the amplitudes in Eqs. (\ref{Eq:jpsipi}), (\ref{Eq:etac}), (\ref{Eq:chic0}) and (\ref{Eq:chic1c}), we can estimate the corresponding partial decay width by,
\begin{eqnarray}
\Gamma(Z_c(3900)^0 \to f) =\frac{1}{3}\frac{1}{8\pi} \frac{|\vec{p}_f|}{m_{Z_c}^2}  \overline{ \left|\mathcal{M}_{Z_c(3900)^0 \to f}\right|^2} ,
\end{eqnarray}
where the overline indicates the sum over the polarization vectors of the vector particles and $\vec{p}_f$ is the momentum of the final state in the $Z_c(3900)^0$ rest frame.

\section{Numerical Results and Discussion}\label{sec3}

\begin{figure}[htb]
  \centering
  \includegraphics[width=80mm]{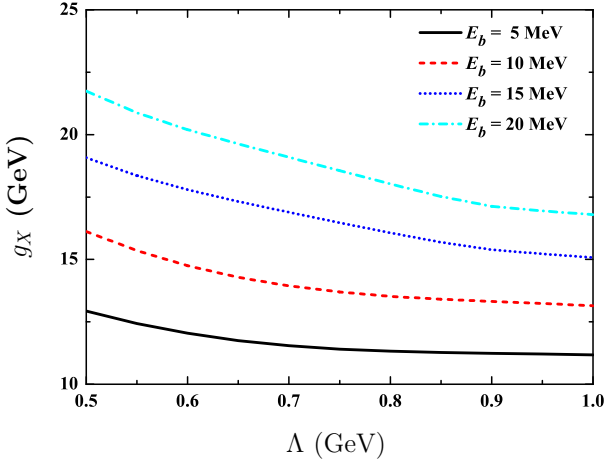}
  \caption{(Color online) The coupling constant $g_X$ depending on the parameter $\Lambda$ and binding energy. }\label{Fig:CP}
\end{figure}

Considering $Z_c(3900)^0$ as a $D \bar{D}^\ast +c.c$ hadronic molecule, the coupling $g_{Z_c}$ can be estimated from the compositeness condition that we presented in the last section. Here, we take several typical values of the binding energy for $Z_c(3900)^0$, which are 5, 10, 15 and 20 MeV. The $\Lambda$ dependence of the coupling constants $g_{Z_c}$ is presented in Fig. \ref{Fig:CP}. The coupling $g_{Z_c}$ monotonously decreases with the increasing of $\Lambda$ and in particular, taking $E_b =5$ MeV as an example, varying $\Lambda$ from 0.5 to 1.0 GeV, we get the value for $g_{Z_c}$ from 12.9 to 11.2 MeV. For a certain $\Lambda$, the coupling increases with the increasing of the binding energy.

\begin{figure}[htb]
  \centering
  \includegraphics[width=80mm]{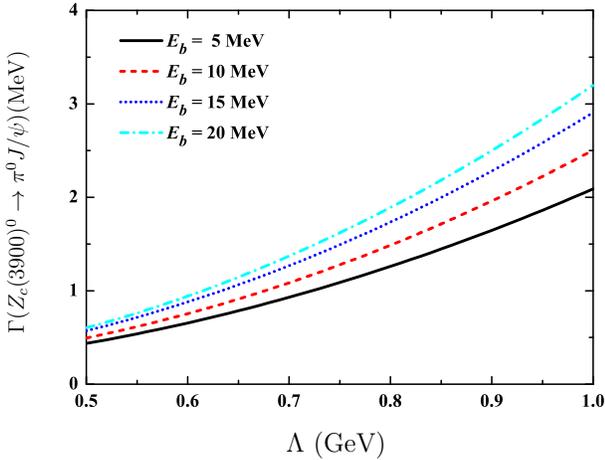}
  \caption{(Color online) The partial decay width of $Z_c(3900)^0 \to \pi^0 J/\psi$ depending on the parameter $\Lambda$ and binding energy. }\label{Fig:psipi}
\end{figure}

The partial decay width of $Z_c(3900)^0 \to \pi^0 J/\psi$ is presented in Fig. \ref{Fig:psipi}. In the discussed $\Lambda$ range, the partial decay width of $Z_c(3900)^0 \to \pi^0 J/\psi$ increases from 0.5 MeV to 2.1 MeV for $E_b=5$ MeV, while for $E_b=20$ MeV, the partial decay width can reach up to 3.2 MeV when  $\Lambda=1.0$ GeV is adopted.  Assuming the charmoniumlike states observed in $Y(4260) \to \pi^\pm (D^\ast D)^\mp$ and $Y(4260) \to \pi^{\pm} (\pi^\mp J/\psi)$ are the same state, one can obtain the ratio of the partial decay width for $J/\psi \pi$ and $D^\ast D$ \cite{Ablikim:2013xfr},
\begin{eqnarray}
\frac{\Gamma(Z_c(3900)^\pm \to (D\bar{D}^\ast)^\pm)}{\Gamma(Z_c(3900)^\pm \to \pi^\pm J/\psi))} = 6.2 \pm 1.1 \pm 2.7,
\end{eqnarray}
With the assumption that these two modes are the dominant decay modes of $Z_c(3900)^\pm$ and using the PDG average value of the $Z_c(3900)$ width $\Gamma_{Z_c(3900)} = 35 \pm 7$ MeV \cite{Agashe:2014kda}, one can estimate the partial decay width $\Gamma(Z_c(3900)^\pm \to \pi^\pm J/\psi)$
to be $2.7 \sim 9.8$ MeV. Considering the isospin symmetry, we can get $\Gamma(Z_c(3900)^0 \to \pi^0 J/\psi \simeq \Gamma(Z_c(3900)^\pm \to \pi^\pm J/\psi)$.  Our estimation of the partial decay width for $Z_c(3900) \to \pi^0 J/\psi $ is consistent with the experimental measurements in the discussed parameter range.

\begin{figure}[htb]
  \centering
  \includegraphics[width=80mm]{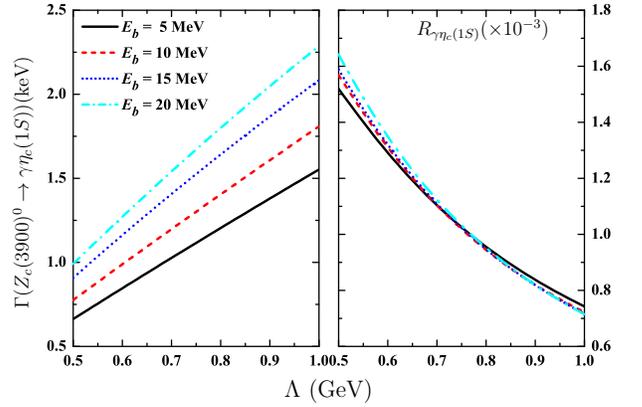}
  \caption{(Color online) The partial decay width of $Z_c(3900)^0 \to \gamma \eta_c(1S)$ (left panel) and the ratio $R_{\gamma \eta_c(1S)}$ (right panel) depending on the parameter $\Lambda$ and binding energy. }\label{Fig:etac1s}
\end{figure}

\begin{figure}[htb]
  \centering
  \includegraphics[width=80mm]{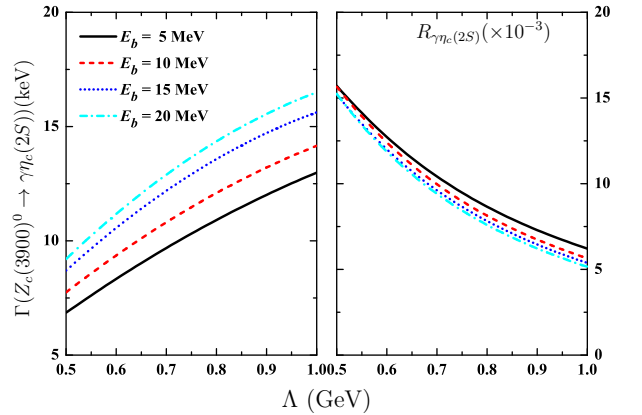}
  \caption{(Color online) The same as Fig. \ref{Fig:etac1s} but for $Z_c(3900)^0 \to \gamma \eta_c(2S)$. }\label{Fig:etac2s}
\end{figure}

We estimate the radiative decay of $Z_c(3900)^0$ to the conventional charmonia in the $D \bar{D}^\ast +c.c$ hadronic molecule scenario. To further reduce the parameter dependence of the radiative decay width, we define the ratio of the radiative partial decay width to the one for ${Z_c(3900)^0 \to \pi^0 J/\psi}$ as,
\begin{eqnarray}
R_{\gamma (c\bar{c})} =\frac{\Gamma(Z_c(3900)^0 \to \gamma (c\bar{c}))}{\Gamma(Z_c(3900)^0 \to \pi^0 J/\psi)}
\end{eqnarray}
where $(c\bar{c})=(\eta_c(1S), \eta_c(2S), \chi_{c0}, \chi_{c1})$. In addition, the cross section of $e^+ e^- \to \pi^0 \pi^0 J/\psi$ and the ratio of $\sigma(e^+ e^- \to \pi^0 Z_c(3900)^0 \to \pi^0 \pi^0 J/\psi)$ to $\sigma(e^+ e^- \to \pi^0 \pi^0 J/\psi)$ have been reported by the BESIII Collaboration \cite{Ablikim:2015tbp}.  Here, we notice that the cross sections for $e^+ e^- \to \pi^0 Z_c(3900)^0 \to \pi^0 \pi^0 J/\psi$ are relative large at $E_{\mathrm{c.m.}} =4.230$ GeV and $E_{\mathrm{c.m.}}=4.260$ GeV, and the cross section for $e^+ e^- \to \pi^0 Z_c(3900)^0 \to J/\psi \pi^0 \pi^0$ is about $9.9 \pm 2.1$ pb and $4.0 \pm 1.2$ pb, respectively.  With the ratio of $R_{\gamma (c\bar{c})}$ and the measured cross section for $e^+ e^- \to \pi^0 Z_c(3900)^0 \to \pi^0 \pi^0 J/\psi$, we can roughly estimate the cross section for $e^+ e^- \to \pi^0 Z_c(3900)^0 \to \pi^0 \gamma (c\bar{c})$ at these two energy points, which would be useful for the experimental search for $Z_c(3900)^0$ in its radiative decay modes.

The $\Lambda$ dependence of the partial decay width for $Z_c(3900)^0 \to \gamma \eta_c(1S)$ and the ratio $R_{\gamma \eta_c(1S)}$ are presented in Fig. \ref{Fig:etac1s}. The partial width for $Z_c(3900)^0 \to \gamma \eta_c(2S)$ is evaluated to be of order keV. In particular, for the case of $E_b=5$ MeV, the partial decay width for $Z_c(3900)^0 \to \gamma \eta_c(1S)$ varies from 0.7 to 1.5 keV when $\Lambda$ increasing from 0.5 to 1 GeV,  while for the case of $E_b=20$ MeV, this partial width increases from 1.0 to 2.3 keV. As for the ratio $R_{\gamma \eta_c(1S)}$, it is of order $10^{-3}$ and very weakly dependent on the binding energy of $Z_c(3900)^0$. This ratio decreases from about $1.6 \times 10^{-3}$ to $0.7 \times 10^{-3}$ as $\Lambda$ increases from 0.5 to 1.0 GeV.  With this ratio, the upper limit of the cross section at 4.220 is 19 fb, which is too small to be observed. The partial decay width for $Z_c(3900)^0 \to \gamma \eta_c(2S)$ and the ratio $R_{\gamma \eta_c(2S)}$ are presented in Fig. \ref{Fig:etac2s}. Comparing to the process $Z_c(3900)^0 \to \gamma \eta_c(1S)$, we find the partial width for $Z_c(3900)^0 \to \gamma \eta_c(2S)$ is nearly 10 times larger than the one for $Z_c(3900)^0 \to \gamma \eta_c(1S)$. The partial width $Z_c(3900)^0 \to \gamma \eta_c(2S)$ can reach up to 17 keV and the ratio $R_{\gamma \eta_c(2S)}$ varies from $15\times 10^{-3}$ to $5 \times  10^{-3}$ with $\Lambda$ increasing from 0.5 to 1 GeV. The cross section for $e^+ e^- \to \pi^0 Z_c(3900)^0 \to \pi^0 \gamma \eta_c(2S)$ can be as high as 0.18 pb at 4.23 GeV.

\begin{figure}[htb]
  \centering
  \includegraphics[width=80mm]{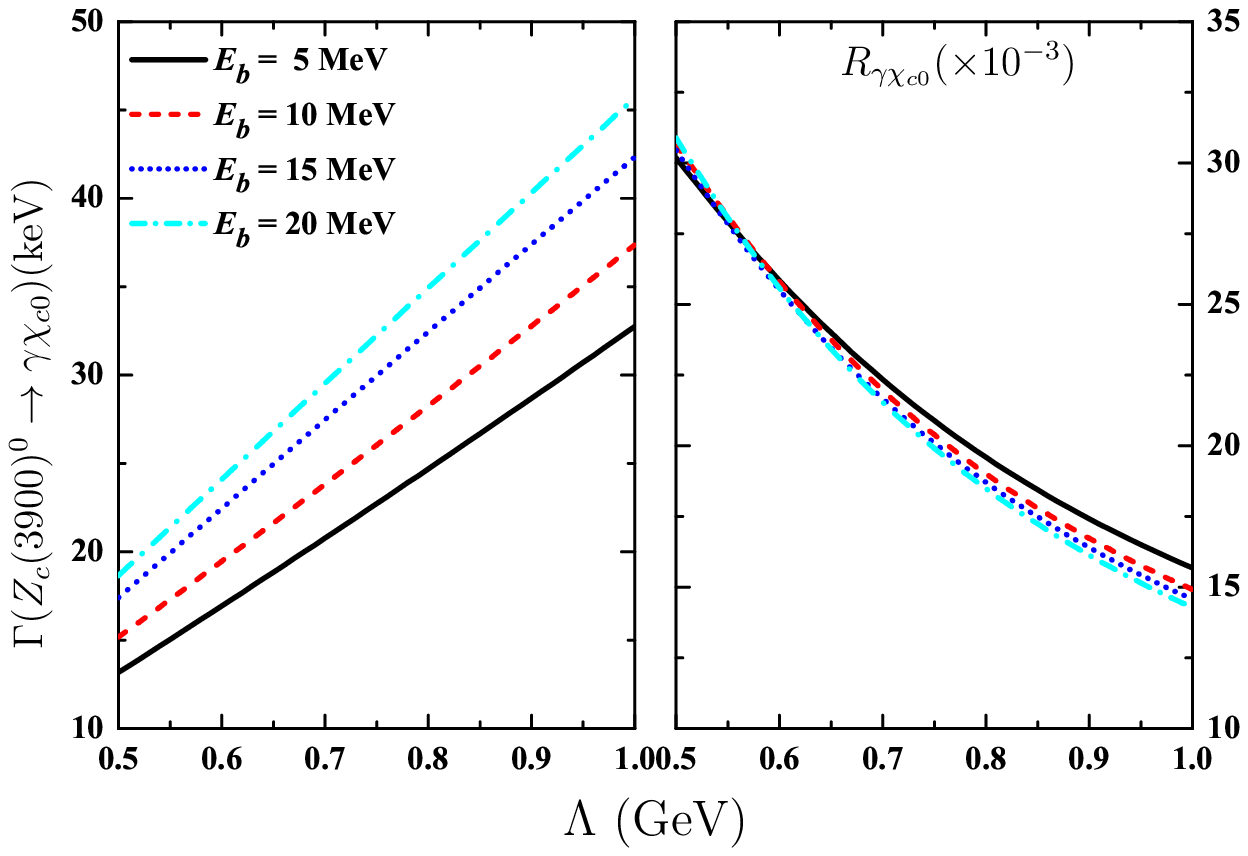}
  \caption{(Color online) The same as Fig. \ref{Fig:etac1s} but for $Z_c(3900)^0 \to \gamma \chi_{c0}$. }\label{Fig:chic0}
\end{figure}

\begin{figure}[htb]
  \centering
  \includegraphics[width=80mm]{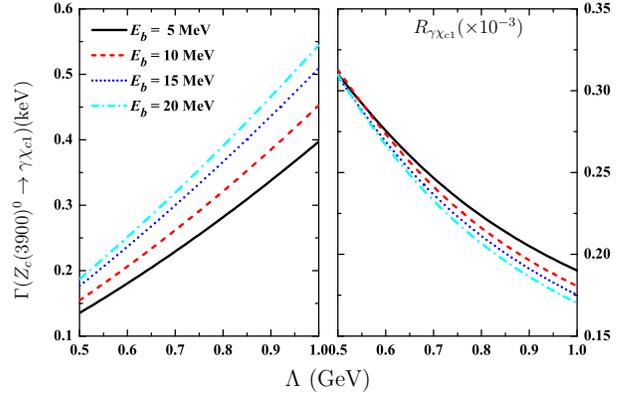}
  \caption{(Color online) The same as Fig. \ref{Fig:etac1s} but for $Z_c(3900)^0 \to \gamma \chi_{c1}$. }\label{Fig:chic1}
\end{figure}

The partial width for $Z_c(3900)^0 \to \gamma \chi_{c0}$ and the ratio $R_{\gamma \chi_{c0}}$ are presented in Fig. \ref{Fig:chic0}. The partial decay width of $Z_c(3900) \to \gamma \chi_{c0}$ is larger than the one for $Z_c(3900)^0 \to \gamma \eta_c(2S)$, it reaches up to 45 keV when $\Lambda=1.0$ GeV and $E_b=20$ MeV. $Z_c(3900)^0$ decays into $\gamma \eta_c/\eta_c(2S)$ via the $S-$ and $D-$wave, but into $\gamma \chi_{c0}$ via the $P-$ and $G(F)-$wave. However, one has to notice that in the $D \bar{D}^\ast$ hadronic molecule scenario, $Z_c(3900)^0$ decays into $\gamma \eta_c/\eta_c(2S)$ via the triangle diagrams in Fig. \ref{Fig:feyn-etac}, where both $D^{(\ast)} D^{(\ast)} \gamma$ and $D^{(\ast)} D^\ast \eta_c$ are $P-$ wave coupling, but in the $Z_c(3900)^0 \to \gamma \chi_{c0}$ process, the $D^{(\ast)} D^{(\ast)}$ couples to $\chi_{c0}$ via $S-$ wave. Thus, it is understandable that the partial width for $Z_c(3900)^0 \to \gamma \chi_{c0} $ is larger than the one for $Z_c(3900) \to \gamma \eta_c(2S)/\eta_c(1S)$. The ratio $R_{\gamma \chi_{c0}}$ can reach up to $3.0 \times 10^{-2}$, which indicates the cross section for $e^+ e^- \to \pi^0 Z_c(3900)^0 \to \pi^0 \gamma \chi_{c0}$ can be up to 0.36 pb at 4.230 GeV and 0.15 pb at 4.260 GeV. The partial width for $Z_c(3900)^0 \to \gamma \chi_{c1}$ and the ratio $R_{\gamma \chi_{c1}}$ are presented in Fig. \ref{Fig:chic1}. The partial width is about two orders smaller than the one for $Z_c(3900)^0 \to \gamma \eta_c(1S)$, which indicates $Z_c(3900)^0 \to \gamma \chi_{c1}$ is very hard to detect.

\section{Summary}\label{sec4}

As the neutral partner of $Z_c(3900)^\pm$, $Z_c(3900)^0$ can radiatively decay into conventional charmonium via annihilating the light quark and antiquark, which provides further opportunity to test different interpretations to $Z_c(3900)$. In present work, we estimated the partial widths for $Z_c(3900)^0$ radiative decays in the hadronic molecule scenario, where $Z_c(3900)^0 $ is treated as a molecule state of $D \bar{D}^\ast +c.c$. Our estimations indicated that in the discussed parameter range, the partial widths for $Z_c(3900)^0 \to \gamma \eta_c(2S)$ and $Z_c(3900)^0 \to \gamma \chi_{c0}$ are of order 10 keV, while the partial widths for $Z_c(3900)^0 \to \gamma \eta_c$ and $Z_c(3900)^0 \to \gamma \chi_{c1}$ are of order keV and less than 1 keV, respectively.

In this calculation we found the partial widths are dependent both on the parameter $\Lambda$ introduced by the correlation function and the binding energy of $Z_c(3900)^0$. To further reduce the parameter dependence of our results, we took $Z_c(3900)^0 \to \pi^0 J/\psi$ as a scale and estimated the ratio of the partial width of the discussed radiative decay to $Z_c(3900)^0 \to \pi^0 J/\psi$. The results showed that the ratios are very weakly dependent on the binding energy of $Z_c(3900)^0$.

With the evaluated ratios $R_{\gamma (c\bar{c})}$ and measured cross section $e^+ e^- \to \pi^0 Z_c(3900)^0 \to \pi^0\pi^0 J/\psi$, we roughly estimated the cross section for $e^+ e^- \to \pi^0 Z_c(3900)^0 \to \pi^0 \gamma (c\bar{c})$, where $(c\bar{c})$ denotes $\eta_c(1S)$, $\eta_c(2S)$, $\chi_{c0}$ or $\chi_{c1}$. Our calculation showed that the cross sections for $\chi_{c1}$ and $\eta_c(1S)$ are too small to be observed, while for $\eta_c(2S)$ and $\chi_{c0}$, the cross section can reach up to 0.18 and 0.36 pb, respectively,  at $E_{\mathrm{cm}}=4.230$ GeV. Experimentally, with current integrated luminosity that BESIII has accumulated at 4.23 GeV, the process $e^+e^- \to \pi^0 Z_c(3900)^0 \to \pi^0 \gamma \chi_{c0}/\eta_c(2S) $ may be searched for without the reconstruction of the $\chi_{c0}/\eta_c(2S)$ signals and with only $\gamma \pi^0$ identified, where $\gamma \pi^0$ recoil mass could be required to be within the $\chi_{c0}/\eta_c(2S)$ mass region to reduce the backgrounds and the $Z_c(3900)$ signals could be searched for in the recoil mass distribution of $\pi^0$. Such research can also be done in the forthcoming Belle II experiment via initial state radiation.

\section*{Acknowledgements}
D.Y.C. would like to thank Cheng-Ping Shen for his careful reading of the manuscript and useful discussion on the experimental possibilities. This project is supported by the National Natural Science Foundation of China under Grant No. 11375240 and No. 11475192.  The fund provided to the Sino-German CRC 110 ``Symmetries and the
Emergence of Structure in QCD" project by the DFG is also appreciated.


\end{document}